\documentclass[reprint,amsmath,amssymb,aps,prl]{revtex4-1}
 
\usepackage{float}
\usepackage{graphicx}
\usepackage{dcolumn}
\usepackage{bm}
\usepackage{hyperref}
\usepackage{siunitx}

\begin{document}

\title{Quasinormal-mode perturbation theory for dissipative and dispersive optomechanics}

\author{Andr\'e G. Primo}
\email{agprimo@ifi.unicamp.br}
\affiliation{Applied Physics Department and Photonics Research Center, University of Campinas, Campinas, SP, Brazil}
\author{Nat\'alia C. Carvalho}
\affiliation{Applied Physics Department and Photonics Research Center, University of Campinas, Campinas, SP, Brazil}
\author{Cau\^e M. Kersul}
\affiliation{Applied Physics Department and Photonics Research Center, University of Campinas, Campinas, SP, Brazil}
\author{Newton C. Frateschi}
\affiliation{Applied Physics Department and Photonics Research Center, University of Campinas, Campinas, SP, Brazil}
\author{Gustavo S. Wiederhecker}
\affiliation{Applied Physics Department and Photonics Research Center, University of Campinas, Campinas, SP, Brazil}
\author{Thiago P. Mayer Alegre}
\email{alegre@unicamp.br}
\affiliation{Applied Physics Department and Photonics Research Center, University of Campinas, Campinas, SP, Brazil}

\date{\today}

\begin{abstract}
Despite the several novel features arising from the dissipative optomechanical coupling, such effect remains vastly unexplored due to the lack of a simple formalism that captures non-Hermiticity in the engineering of optomechanical systems. In this Letter, we show that quasinormal-mode-based perturbation theory is capable of correctly predicting both dispersive and dissipative optomechanical couplings. We validate our model through simulations and also by comparison with experimental results reported in the literature. Finally, we apply this formalism to plasmonic systems, used for molecular optomechanics, where strong dissipative coupling signatures in the amplification of vibrational modes could be observed.
\end{abstract}

\maketitle

\begin{figure}[ht!]
\includegraphics[width = 8.2cm]{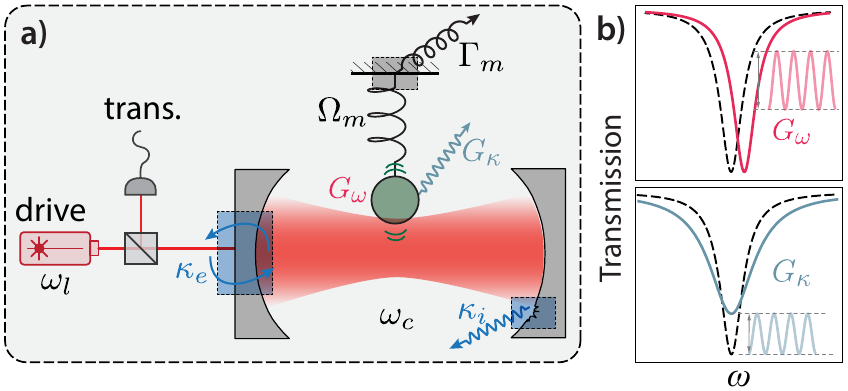}
\caption{\small{\textbf{(a)} Optomechanical coupling scheme. The cavity mode with frequency $\omega_c$ interacts with a mechanical mode of frequency $\Omega_m$ and damping rate $\Gamma_m$. The interaction is mediated by dissipative ($G_\kappa$) and dispersive ($G_\omega$) optomechanical couplings. \textbf{(b)} $G_\omega$ and $G_\kappa$ modulate the number of photons in the cavity mode.}}
\label{fig:1}
\end{figure}

\textit{Introduction.}--- Cavity optomechanics has been a prolific field of research in the past decades~\cite{Aspelmeyer2014CavityOptomechanics, Wiederhecker2019BrillouinStructures}, with applications in quantum information processing~\cite{Stannigel2012OptomechanicalPhonons}, microwave-to-optical signal conversion~\cite{Jiang2020EfficientFrequency,Shao2019Microwave-to-opticalResonators}, sensing and precision measurement~\cite{Yang2015SimpleSensing,Zhang2012SynchronizationLight}, and as platform for fundamental physics tests~\cite{Marinkovic2018OptomechanicalTest,Delic2020CoolingState,Chan2011LaserState}. Dispersively coupled optical and acoustic modes, with frequency pulling 
$G_\omega~=~-\frac{d\omega}{dx}$, have been at the core of most developments in this field~\cite{Baumberg2019ExtremeGaps,Benevides2017Ultrahigh-QFoundry,Leijssen2015StrongNanobeam,Lombardi2018PulsedBond-Breaking}.
However, a complete description of the optomechanical interaction must take into account a dissipative coupling $G_\kappa = \frac{d\kappa}{dx}$, where a mechanical displacement $x$ further modulates the optical mode's linewidth, which can be of external nature ($\kappa_e$), associated to the coupling to a coherent excitation channel, and/or internal ($\kappa_i$), containing the losses through all other channels, (Figs.~\ref{fig:1} \textbf{a)} and \textbf{b)}). Various novel features have been predicted using dissipative coupling, i.e. cooling in the bad-cavity limit regime ($\kappa/2\gg\Omega_m$)~\cite{Weiss2013QuantumSystems, Princepe2018Self-SustainedDevices} and quantum-limited position measurement~\cite{Elste2009QuantumNanomechanics}, nonetheless, experimental demonstrations are still scarce~\cite{Wu2014DissipativeSensor,Barnard2019Real-timeNanotube} and lack a theoretical framework to engineer devices with strong dissipative coupling. 

Although other formulations are possible, optomechanical interactions are typically modeled using perturbation theory, providing insight into phenomena through simple mathematical expressions which ease their understanding and engineering. The interplay between optical and mechanical modes results from deformation-dependent changes in the permittivity $\Delta \bm\varepsilon(x)$. In mesoscopic mechanical resonators (dimensions on the order of the optical wavelength), $\Delta \bm\varepsilon(x)$ arises from two main mechanisms: moving boundaries (MB)~\cite{Johnson2002PerturbationBoundaries} and photoelasticity (PE)~\cite{Balram2014MovingResonators}. In the microscopic regime, e.g. molecular optomechanics, molecules are treated as point dipoles interacting with plasmonic modes and the optomechanical coupling arises from their mechanical displacement-dependent polarizability $\alpha(x)$.

In this Letter, we apply a non-Hermitian perturbation theory based on quasinormal-modes (QNM) or modes of an open resonator~\cite{Muljarov2010Brillouin-WignerSystems, Weiss2016FromSensing,Yang2015SimpleSensing,Cognee:19}, to simultaneously evaluate $G_\kappa$ and $G_\omega$ in arbitrary geometries undergoing small deformations. We use finite-element method (FEM) to demonstrate the accuracy of our method in two conceptual devices chosen to illustrate distinct aspects of this formulation: a ring resonator interacting with an absorptive element and a split-beam nanocavity, experimentally reported in Ref.~\cite{Wu2014DissipativeSensor}; in these examples, we neglect the presence of waveguides in the calculations and therefore focus on the internal
linewidth perturbation case only, i.e. $G_\kappa = d\kappa_i/dx$. Lastly, we apply our formalism to a nanoparticle-on-a-mirror (NPoM)~\cite{Daniels2005NanoparticleScattering, Mubeen2012PlasmonicOxide} scheme, where we show that plasmonic nanocavities used in the context of surface or tip-enhanced Raman scattering (SERS or TERS) could naturally display large dissipative couplings which should be carefully evaluated when considering such platform for quantum optomechanics experiments.

Previous perturbation theory formulations for the dispersive optomechanical coupling use a Hermitian (or lossless) treatment of electromagnetic fields, even in cases where dissipation is present. In Hermitian systems, optical normal-modes (NM) $\vec{E}_{m}$ arise and are power-orthogonal~\cite{Snyder1983a}, i.e. $ \int \vec{E}^*_m \cdot \bm\varepsilon \vec{E}_n dV = C_{n,m}\delta_{n,m}$, where $\bm\varepsilon$ is the permittivity tensor of the medium, $\delta_{n,m}$ the Kronecker delta and $C_{n,m}$ a normalization constant with dimensions of energy. This orthogonality relation is one of the \textit{foundations} on which the optomechanical coupling is derived. Lossy structures, however, are described by non-Hermitian operators, and rigorously disrupt power-orthogonality, hindering the limits of applicability of this formulation. Here, we treat optical modes as inherently lossy and re-derive the optomechanical coupling expressions by introducing corrected orthogonality relations through QNM bi-orthogonality~\cite{Moiseyev2011Non-hermitianMechanics, Kristensen2020ModelingModes, Pick2017GeneralPoints,Lalanne2018LightResonances,Miri2019,Lai2019ObservationEffect, Hodaei2017EnhancedPoints}. Recent work used this formalism to describe modifications on the optical response of lossy resonators due to designed static boundary deformations~\cite{Yan2020ShapeTheory}. Here, we provide the link with optomechanics, where time-dependent deformations arise from mechanical normal modes inherent to the system, and thus a coupled description of the elastic and optical responses is required. By generalizing optomechanical coupling using the notion of QNMs, we provide a straightforward way of accurately simultaneously computing its dispersive and dissipative components.

Bi-orthogonality relations are derived between left $\vec{E}_n^L$ and right $\vec{E}_n^R$ eigenvectors of a generalized eigenvalue problem~\cite{Seyranian2003MultiparameterApplications,Siegman1986Lasers}; in non-dispersive media they read: $\int \vec{E}_n^L\cdot \bm\varepsilon \vec{E}_m^R dV = C_{n,m}\delta_{n,m}$,
where a reciprocal medium is assumed, i.e. $\bm \varepsilon =\bm \varepsilon^T$ with ``$T$" denoting transposition. Since left and right eigenmodes \textit{are not complex conjugate pairs}, this expression is fundamentally different from power-orthogonality and generally yields complex values. In section S1 of the Supplemental Material $\vec{E}_n^L$ and $\vec{E}_n^R$ are shown to be counter-propagating degenerate mode pairs in traveling-wave resonators, or, exactly equal in standing-wave resonators. Although QNMs have not been proven to span a complete basis, QNM-based perturbation theories display remarkable accuracy when perturbations take place internally or in close proximity to the resonator~\cite{Muljarov2010Brillouin-WignerSystems, Weiss2016FromSensing, Yan2020ShapeTheory} and also for isotropic and homogeneous external perturbations~\cite{Both2019First-orderResonators}. Here, we circumvent this difficulty by considering a finite computational domain bounded by perfectly matched layers (PML), in which completeness is recovered~\cite{Yan2018RigorousNanoresonators}. Due to our choice, all volume integrals stated must also be performed in the PML region, with appropriate coordinate transformations, as discussed in detail in Ref.~\cite{Sauvan2013TheoryResonators}.

\begin{figure*}[t!]
\centering
\includegraphics[width = 16cm]{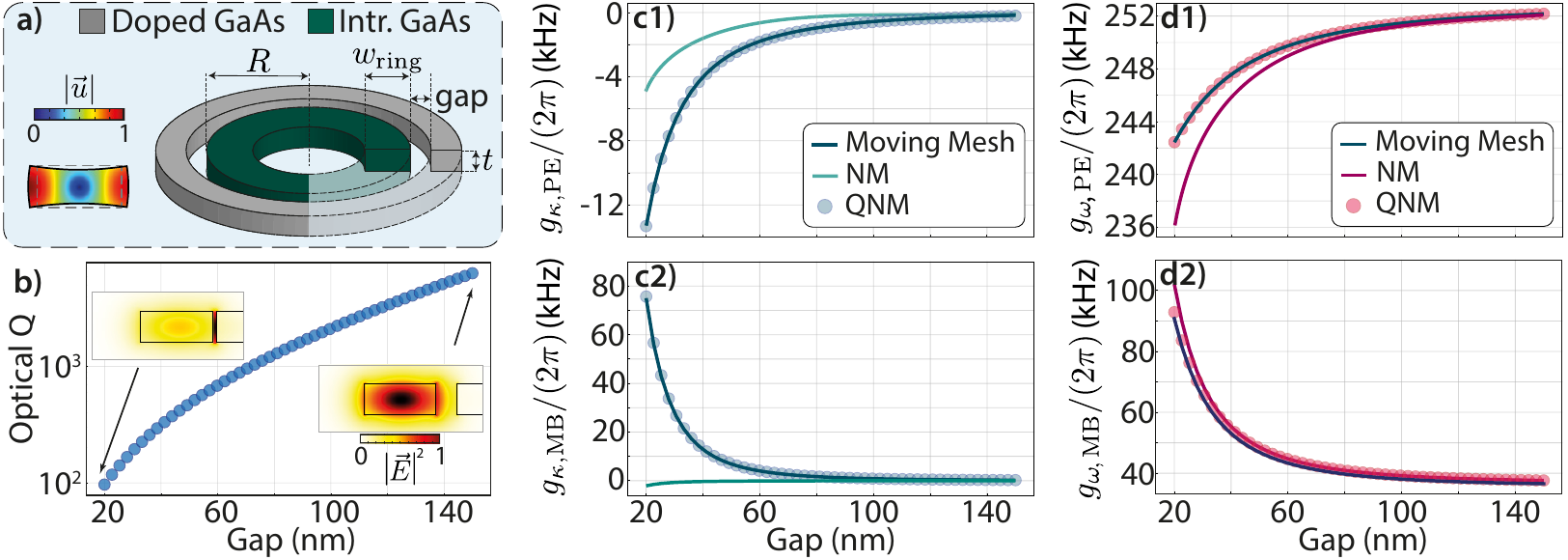}
\caption{\small{\textbf{(a)} Concentric doped and intrinsic GaAs rings with relevant geometric parameters ($R=\SI{3.5}{\micro\metre}$, $w_\mathrm{ring}=\SI{500}{\nano\metre}$ and $t=\SI{220}{\nano\metre}$) and mechanical breathing mode profile. \textbf{(b)} Q-factor as a function of gap between rings. \textbf{(c1)} Phototelastic (PE) and \textbf{(c2)} moving-boundary (MB) contributions for the zero-point dissipative coupling calculated using the normal-mode (NM) and quasinormal-mode (QNM) perturbation theories, compared with full-wave calculations (Moving Mesh). \textbf{(d1)} PE and \textbf{(d2)} MB contributions for the zero-point dispersive coupling.}}
\label{fig:2}
\end{figure*}

\textit{Numerical modeling.}--- To validate our approach, we consider a model composed of two concentric $\SI{250}{\nm}$ thick GaAs rings, as shown in Fig.~\ref{fig:2} \textbf{a)}. The (inner) ring resonator, which supports traveling-wave type optical modes, is undoped ($\varepsilon/\varepsilon_0=n
^2$), where $n = 3.46$ and $\varepsilon_0$ the vacuum permittivity, while its width and radius are $w_\text{ring} = \SI{500}{\nm}$ and $R = \SI{3.5}{\micro\metre}$, respectively. The external ring is heavily doped and its dielectric constant is given by $\varepsilon/\varepsilon_0 =  n^2-j\sigma/(\varepsilon_0\omega)$, with conductivity $\sigma = \SI{5e4}{\siemens/\meter}$, similar to values achieved with ion implantation in semiconductor substrates; its width is chosen to  suppress strong reflections at the perfect electric conductor boundary that limits the simulation domain. This example has the advantage of admitting complex eigenvalues regardless of radiation losses, constituting a relevant case for absorption-induced dissipative coupling \cite{Barnard2019Real-timeNanotube, Princepe2018Self-SustainedDevices}.

The inner ring's mechanical breathing mode ($\Omega_m/(2\pi) \approx \SI{4.1}{\GHz}$) is considered, while the external doped structure is kept fixed. For a given gap between the two rings, the optomechanical coupling was evaluated using both QNM-based (bi-orthogonality) and the NM (power-orthogonality) approaches. Their distance was varied to encompass both high and low optical-Q regimes, as shown in Fig.~\ref{fig:2} \textbf{b)}, allowing us to probe the domains of validity of both methods. The calculated optomechanical coupling rates were then cross-checked against exact simulations that incorporate both MB and PE effects. For that purpose, the optomechanical ring's geometry is deformed following the mechanical mode profile using COMSOL\textsuperscript{\textregistered}'s \textit{Moving Mesh} module, including the refractive index spatial dependency due to photoelasticity in the permittivity tensor. The derivatives of the real and imaginary parts of the optical frequency relative to the mechanical deformation amplitude provide -- within numerical precision -- an accurate calculation of the optomechanical couplings.

The QNM perturbative calculations for moving boundary $G_\text{MB}$ and photoelastic $G_\text{PE}$ optomechanical couplings (in units of \SI{}{\Hz/\meter}) are given by (see S2 in the Supplemental Material for details):
\begin{subequations}
\begin{align}
    G_\text{MB} &= -\frac{\omega_{0}}{2}\frac{\int d\vec{A}{\cdot}\vec{u}\left( \vec{E}^{L}_{\parallel}{\cdot}\Delta\bm\varepsilon\vec E^{R}_{\parallel}-\vec{D}^{L}_{\perp}{\cdot}\Delta(\bm\varepsilon^{-1})\vec D^{R}_{\perp} \right)}{\int dV \vec{E}^L{\cdot}\big(\bm\varepsilon(\vec{r},\omega_0)+\frac{\omega_{0}}{2}\frac{\partial \bm \varepsilon(\vec{r},\omega_{0})}{\partial \omega}\big)\vec E^{R}},\label{eq:pert_theory1}  \\
    G_\text{PE} &= \frac{\omega_{0}}{2}\frac{ \int dV \bm\varepsilon^2 \left( \vec{E}^{L}{\cdot} \bm p \mkern1mu{:} \bm S \vec E^{R} \right)}{\int dV \vec{E}^L{\cdot}\big(\bm\varepsilon(\vec{r},\omega_0)+\frac{\omega_{0}}{2}\frac{\partial \bm \varepsilon(\vec{r},\omega_{0})}{\partial \omega}\big)\vec E^{R}},
    \label{eq:pert_theory2}    
\end{align}
\end{subequations}
where $\vec{D}^{R(L)}$ denotes the electric displacement field related to the right (left) electric fields,  $\omega_{0}$ is the unperturbed (complex) frequency of the optical resonator, $\vec{u}$ is the unit-normalized mechanical displacement associated with the elastic strain $\bm S$, $\bm p$ is photoelastic tensor, $\perp$ and $\parallel$ denote the perpendicular and parallel field components at the mechanical resonator's surface, and  $\Delta\bm\varepsilon=\bm\varepsilon_1-\bm\varepsilon_2$, $\Delta(\bm\varepsilon^{-1})=(\bm\varepsilon_1)^{-1}-(\bm\varepsilon_2)^{-1}$, are related to the permittivities of the guiding ($\bm\varepsilon_1$) and surrounding ($\bm\varepsilon_2$) media. Due to material dispersion, $\bm \varepsilon$ is taken to be spatially and frequency dependent (see Supplemental S2). The couplings defined in Eqs. \ref{eq:pert_theory1}, \ref{eq:pert_theory2} are complex numbers and can be directly associated with their dispersive and dissipative components via $G_{\omega,\text{MB(PE)}} = - \Re(G_{\text{MB(PE)}})$ and $G_{\kappa,\text{MB(PE)}} = 2 \Im(G_{\text{MB(PE)}})$, where $\Re$ and $\Im$ denote real and imaginary parts of the quantity in parenthesis.

The zero-point dissipative optomechanical coupling rates ($g_\kappa = G_\kappa x_\text{zpf}$, where $x_\text{zpf}$ is the zero-point fluctuation of the mechanical mode), are displayed in Fig.~\ref{fig:2} \textbf{c)}. Even in high-Q regimes the NM approximation fails to predict $g_\kappa$ accurately, while the discrepancy grows at smaller gaps.  On the dispersive side ($g_\omega = G_\omega x_\text{zpf}$), shown in Fig.~\ref{fig:2}~\textbf{d)}, differences in the predictions of the NM and QNM-based analysis arise for tighter gaps, indicating that in strongly dissipative systems (as in the case of plasmonic resonators), our approach should be used to correctly calculate $g_\omega$. However, both NM and QNM calculations approach each other for the dispersive coupling at high-Q.

\begin{figure}[t]
\includegraphics[width = 8.2cm]{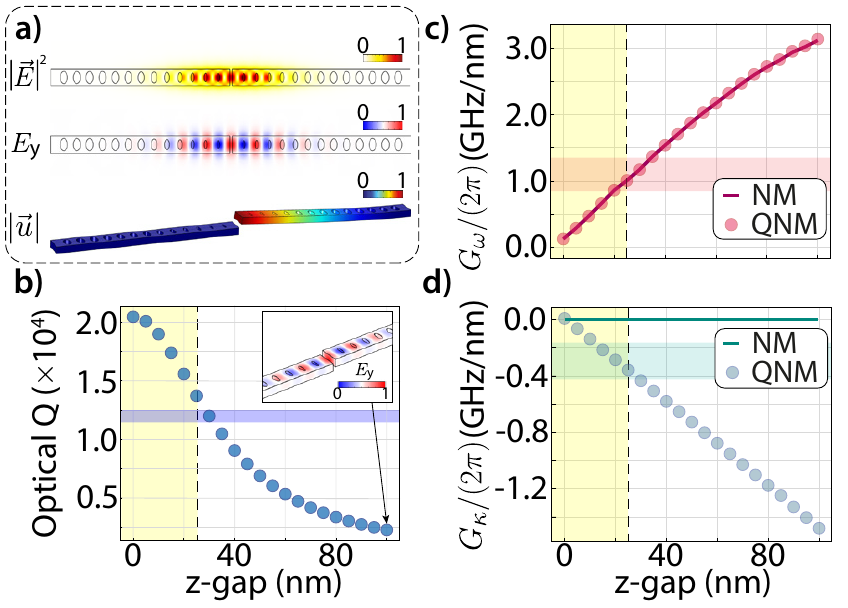}
\caption{\small{\textbf{(a)} Optical and mechanical modes of the split-beam nanocavity. \textbf{(b)} Optical Q-factor as a function of the offset between the two cantilevers (z-gap). Inset: optical mode for $\SI{100}{\nano\metre}$ z-gap. \textbf{(c)} Dispersive coupling $G_\omega$ and \textbf{(d)} dissipative coupling $G_\kappa$ as functions of z-gap calculated using the normal-mode (NM) and quasinormal-mode (QNM) perturbation theories. The yellow vertical and light colored horizontal strips represent respectively the uncertainty in z-gap and measured values as reported in Ref.~\cite{Wu2014DissipativeSensor}.}}
\label{fig:3}
\end{figure}

\textit{Integrated photonic resonators.} --- We turn our attention to experimental results where an appreciable dissipative coupling is observed in order to put the present formalism to test. To the best of our knowledge, the split-beam nanocavity of Ref.~\cite{Wu2014DissipativeSensor} displays the largest $g_\kappa$ reported to date in integrated devices. This 1-D photonic crystal structure is composed by two independent mechanical cantilevers separated by a gap which also serves as defect for high-Q optical confinement. The optical and mechanical modes considered here are displayed in Fig.~\ref{fig:3}~\textbf{a)}. Differences in the anchoring of the two cantilevers break the vertical symmetry of the system, yielding an offset ($z$-gap) between them, degrading the optical Q-factors as in Fig.~\ref{fig:3} \textbf{b)}. The reported uncertainty in $z$-gap is $<\SI{25}{\nano\metre}$ and is represented by the yellow shaded areas in Fig.~\ref{fig:3}, while experimental values for the optical quality factors are represented by blue strip-covered region. The Q-factors were calculated by introducing a carefully implemented PML at the boundaries of the simulation domain. Values obtained for z-gap $= \SI{25}{\nano\metre}$ ($Q_\mathrm{sim} \approx 14k$) agree within 15\% to the experimental values ($Q_\mathrm{exp} \approx 12k$); such small discrepancies are possibly due to design differences between the fabricated and simulated devices. The tapered fiber used in the experiment as an excitation channel is not considered in our simulations and the external dissipative coupling $g_{\kappa_e}$ was neglected. This approximation does not impact our results, where only the measured and simulated internal dissipative couplings are compared.

The optomechanical couplings were evaluated using QNM and NM-based perturbation theories, while accounting for both PE and MB contributions. In the NM calculations, the PML domain is excluded from all integrations, such that the dissipative coupling arises from the imaginary part of the eigenvalue $\omega_0$ alone (Supplemental Material S2). The measured dispersive coupling $G_\omega$ (light red strip) agrees well with both predictions, as shown in Fig.~\ref{fig:3}~\textbf{c)}. This agrees with the discussion for the ring resonator, where both formalisms accurately describe $G_\omega$ for low optical dissipation. $G_\kappa$ is displayed in Fig.~\ref{fig:3}~\textbf{d)}; in this example, the two approaches yield strikingly different predictions, while agreement with measured values (cyan strip) is only obtained using QNM-based theory, corroborating our previous model where the dissipative coupling is only correctly captured by the QNM perturbation theory, regardless of the Q-factors considered. This analysis reinforces the utility of QNMs as a new tool for engineering non-Hermitian effects in optomechanical systems.

\begin{figure*}[t]
\centering
\includegraphics[width = 15.0cm]{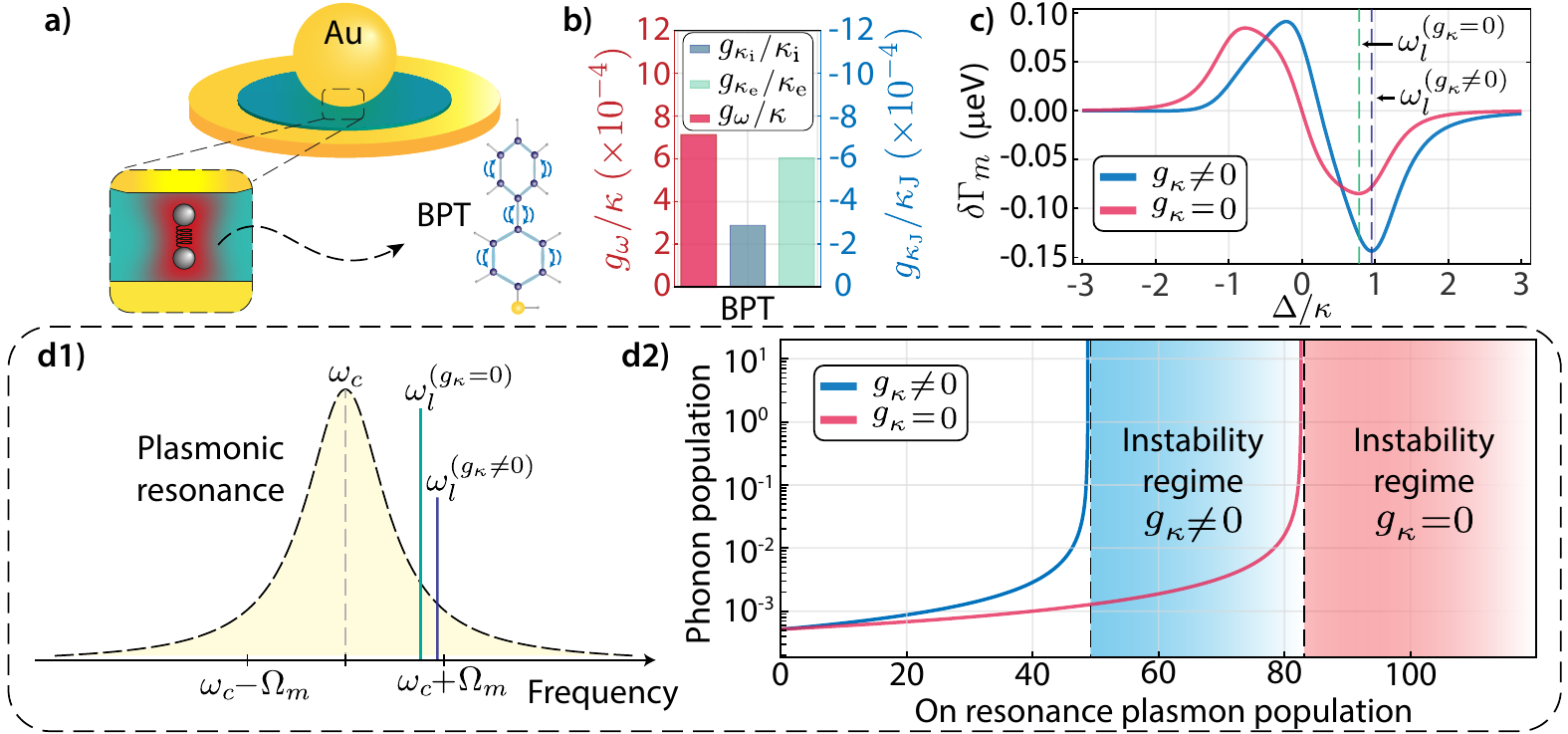}
\caption{\small{\textbf{(a)} Nanoparticle-on-a-mirror (NPoM) scheme and molecular vibrational mode considered for Biphenyl-4-thiol (BPT) according to Ref~\cite{Baumberg2019ExtremeGaps}. \textbf{(b)} Dimensionless dispersive/dissipative coupling strengths $g_\omega/\kappa$, $g_{\kappa_\text{e}}/\kappa_\text{e}$, and $g_{\kappa_\text{i}}/\kappa_\text{i}$ for the NPoM mode interacting with a single BPT molecule. In our system, $\kappa_\text{abs}/(2\pi) \approx \SI{19.3}{\tera\hertz}$, $\kappa_\text{e}/(2\pi) \approx \SI{4.7}{\tera\hertz}$, $\kappa_\text{rad}/(2\pi) \approx \SI{22.7}{\tera\hertz}$, $g_{\kappa_\text{abs}}/(2\pi) \approx \SI{1.6}{\giga\hertz}$, $g_{\kappa_\text{rad}}/(2\pi) \approx -\SI{13.8}{\giga\hertz}$, and $g_{\kappa_\text{e}}/(2\pi) \approx -\SI{2.8}{\giga\hertz}$. All calculations were performed using $\kappa_e/\kappa =  10\%$~\cite{Baumberg2019ExtremeGaps}. \textbf{(c)} Mechanical linewidth  variation $\delta\Gamma_m$ due to dynamical backaction. Graphs are plotted in the presence ($g_\kappa\neq 0$) and absence of ($g_\kappa = 0$) of dissipative coupling. \textbf{(d1)} Scaled schematic plasmonic resonance for pumps $\omega_l^{(g_\kappa=0)}$ and $\omega_l^{(g_\kappa\neq0)}$, optimally chosen for amplifying mechanical motion in the cases with and without $g_\kappa$, as shown by the vertical dashed lines in \textbf{(c)}. \textbf{(d2)} Different instability regimes obtained for molecular collective oscillations with $100$ BPT molecules interacting with a single plasmonic mode. }}
\label{fig:4}
\end{figure*}
\textit{Plasmonic resonators.}  Although the phenomena of SERS or TERS --- where Raman scattering from a molecule is drastically enhanced when near a metallic structure --- has been known for decades, its connection with cavity optomechanics was only pointed out recently~\cite{Roelli2016MolecularScatteringb, Schmidt2016QuantumCavities}. Some important features observed in experiments were qualitatively captured by the introduction of dynamical backaction. Despite the low optical $Q$s found in such plasmonic structures, there are currently no reports that formulate their optomechanical coupling on the basis of QNMs and the role played by the dissipative coupling has been ignored. Qualitatively, the dissipative coupling in SERS experiments arises from light radiated by the molecule interfering constructively (destructively) with light in the QNM's radiation field or in the absorbing metal, leading to an increase (decrease) in the losses of the system~\cite{Cognee:19}. 

The molecular optomechanical coupling is calculated considering a vibrational point-dipole optimally aligned with the interacting electric field~\cite{Lombardi2018PulsedBond-Breaking,Roelli2016MolecularScatteringb}. A derivation of the QNM-based optomechanical coupling for this case is given in S2 of the Supplemental Material and resembles the one given in Ref.~\cite{Roelli2016MolecularScatteringb}, albeit using a different expression for the modal volume.

We demonstrate the relevance of the dissipative coupling in plasmonic resonators through FEM simulations of a gold NPoM scheme depicted in Fig.~\ref{fig:4}~\textbf{a)}. The sphere (radius $\SI{40}{\nano\metre}$) and mirror are spaced by a $\SI{1.3}{\nano\metre}$ wide dielectric gap  ($\varepsilon/\varepsilon_0 = 2.1$). The plasmonic modes supported are assumed to interact with Biphenyl-4-thiol molecules (BPT, frequency $\Omega_m/(2\pi) = \SI{47.52}{\tera\hertz}$ ($\SI{196.5}{\milli\electronvolt}$), mechanical damping rate $\Gamma_m/(2\pi) = \SI{16.93}{\giga\hertz}$ ($\SI{0.07}{\milli\electronvolt}$)). A longitudinal dipole plasmonic mode ($Q\approx 9$, $\omega_c/(2\pi) \approx \SI{418}{\tera\hertz}$) is used. QNMs were computed using a Drude-Lorentz model for the gold structures along with a PML implementation (section S3 in the Supplemental Material). Nonclassical correlations due to electronic length scales~\cite{Yang2019AElectromagnetism} were neglected.

The dimensionless optomechanical coupling strengths $g_\omega/\kappa$, $g_{\kappa_\text{e}}/\kappa_\text{e}$, and $g_{\kappa_\text{i}}/\kappa_\text{i}$ were evaluated in Fig.~\ref{fig:4} \textbf{b)}, displaying strong dissipative coupling. Plasmonic resonators are typically driven using coherent free-space illumination, i.e. coupling to its radiation fields, therefore the resonator's radiative decay is broken into external ($\kappa_\text{e}$) and internal ($\kappa_\text{rad}$) contributions; $\kappa_\text{rad}$ along with absorptive losses $\kappa_\text{abs}$ compose the resonator's total internal losses ($\kappa_\text{i} = \kappa_\text{rad}+ \kappa_\text{abs}$).  This distinction is important due to different optomechanical responses arising if a dissipative channel is coherently or incoherently driven.   Since only one QNM is excited~\cite{ Lombardi2018PulsedBond-Breaking}, perturbations on the molecule's polarizability do not affect appreciably the directionality of emitted radiation~\cite{Sauvan2013TheoryResonators,Kongsuwan2020PlasmonicRadiation}; as a result  $g_{\kappa_\text{e}}/g_{\kappa_\text{rad}} \approx  \kappa_\text{e}/\kappa_\text{rad}$, as shown in section S4 of the Supplemental Material. We stress that such relation is not valid if there is a spectral overlap between distinct QNMs~\cite{Sauvan2013TheoryResonators}. Moreover, coupling arising from absorption $g_{\kappa_\text{abs}}$ can be estimated, along with $\kappa_\text{abs}$, through simulations in the absence of PMLs.

In Fig.~\ref{fig:4} \textbf{c)} we used quantum noise calculations -- fully encompassing the role played by dissipative coupling~\cite{Weiss2013QuantumSystems,Weiss2013Strong-couplingSystems} -- to evaluate the optical cooling/amplification of a single molecule~(see S4 in the Supplemental Material) coupled to a QNM, as a function of the detuning between the driving laser and plasmonic resonance $\Delta/\kappa$. An input power from an external laser yielding an average occupation of a single plasmon at resonance ($\Delta = 0$) was considered. In our case, the presence of $g_\kappa$, with its external and internal components calculated using the aforementioned approximations, significantly enhances optical amplification since the relative phase between $g_\kappa$ and $g_\omega$ leads to a constructive interference between dissipative and dispersive amplification processes. A blue-shift in the maximum Raman enhancement frequency when compared to purely dispersive results is also predicted, a feature that has been observed experimentally and is reported in Refs.~\cite{Roelli2016MolecularScatteringb,Zhu2014QuantumScattering}. Notably, our results depend explicitly on the value of $\kappa_e$ only through the plasmon population, therefore our conclusions are quantitatively valid even if $\kappa_e$ is modified. 

Finally, we consider $N = \SI{100}{}$ BPT molecules coupled to a plasmonic mode as in Ref.~\cite{Lombardi2018PulsedBond-Breaking}. This setup gives rise to molecular coherent collective self-sustaining oscillations through an effectively enhanced optomechanical coupling factor $g_\text{eff} = \sqrt{N} g$. We choose pump laser frequencies that yield maximal amplification in the cases $\omega_l^{(g_\kappa\neq0)}$ and  $\omega_l^{(g_\kappa=0)}$, as schematically shown at scale in Fig.~\ref{fig:4}~\textbf{d1)}. In Fig.~\ref{fig:4}~\textbf{d2)} we observe that the onset of instability is verified to occur at $\approx 40\%$ lower driving powers if dissipative coupling is considered, in agreement with results Ref.~\cite{Lombardi2018PulsedBond-Breaking} where instability is verified to occur at lower driving powers than predictions using $g_\omega$ alone.

\textit{Conclusion.} --- We have shown that QNMs can be used to capture effects that arise from non-Hermiticity in optomechanical systems. This formalism provides a powerful perturbation theory tool and insight into the engineering of such effects in integrated photonics, opening possibilities for strong dissipative coupling in ultra-high-Q devices ($Q>10^6$), where optomechanical transduction through this mechanism is enhanced. This work also brings in the relevance of the phenomenon for low-Q plasmonic devices, where dynamical backaction is greatly affected by the presence of dissipation. This interplay is potentially important for a proper understanding of the Raman spectrum of molecules and for prototyping the next generation of plasmonic devices. Lastly, the effects of such coupling in optomechanical systems operating in the PT-symmetric regime or near an exceptional point remains vastly unexplored. Since those rely on intrinsically non-Hermitian physics, the dissipative coupling may be particularly important and thus a source for a plethora of interesting phenomena.

\textit{Note.} --- FEM and scripts files for generating each figure are available at the ZENODO repository (10.5281/zenodo.3981647)~\cite{zenodo_nonhermitian}. 

\begin{acknowledgments}
This work was supported by S\~ao Paulo Research Foundation (FAPESP) through grants 2019/09738-9, 2017/19770-1, 2017/14920-5, 2020/06348-2, 2018/15580-6, 2018/15577-5, 2018/25339-4, Coordena{\c c}\~ao de Aperfei\c coamento de Pessoal de N{\'i}vel Superior - Brasil (CAPES) (Finance Code 001), Financiadora de Estudos e Projetos, and Conselho Nacional de Desenvolvimento Científico e Tecnológico (CNPq) through grant 425338/2018-5.
\end{acknowledgments}

\end{document}


\title{Supplemental Material: Quasinormal-mode perturbation theory for dissipative and dispersive optomechanics}

\author{Andr\'e G. Primo}
\email{agprimo@ifi.unicamp.br}
\affiliation{Applied Physics Department and Photonics Research Center, University of Campinas, Campinas, SP, Brazil}
\author{Nat\'alia C. Carvalho}
\affiliation{Applied Physics Department and Photonics Research Center, University of Campinas, Campinas, SP, Brazil}
\author{Cau\^e M. Kersul}
\affiliation{Applied Physics Department and Photonics Research Center, University of Campinas, Campinas, SP, Brazil}
\author{Newton C. Frateschi}
\affiliation{Applied Physics Department and Photonics Research Center, University of Campinas, Campinas, SP, Brazil}
\author{Gustavo S. Wiederhecker}
\affiliation{Applied Physics Department and Photonics Research Center, University of Campinas, Campinas, SP, Brazil}
\author{Thiago P. Mayer Alegre}
\email{alegre@unicamp.br}
\affiliation{Applied Physics Department and Photonics Research Center, University of Campinas, Campinas, SP, Brazil}

\date{\today}


\maketitle

\section{Electromagnetic eigenvalue problem}

The eigenmodes of an optical system are calculated from the generalized eigenvalue equation for electromagnetism:

\begin{equation}
    \nabla \times \nabla \times \vec{E}(\vec{r})=\left(\frac{\omega}{c}\right)^{2} \bm{\varepsilon}(\vec{r}) \vec{E}(\vec{r}).
    \label{eigenvalue}
\end{equation}

As pointed out in the main text, the optomechanical interaction is perturbatively calculated by introducing a change in the dielectric constant $\Delta\bm{\varepsilon}$ which is calculated from the mechanical modes profiles. However, the typical description of such phenomenon is based on very particular properties of the operators $\nabla \times \nabla \times$ and $\bm{\varepsilon}$, namely: they are both taken as Hermitian and the latter is positive semi-definite. A series of theorems in linear algebra prove that in this case, the solutions to Eq.~\ref{eigenvalue} are a complete, power-orthogonal set, and its eigenvalues are real.

In the case of dissipative systems, this is not necessarily true:  frequencies become complex and operators are no longer Hermitian. Specifically, we now allow a complex dielectric constant $\varepsilon$ which violate some hyphothesis of the theorems that applied earlier. Consequently, eigenfunctions are no longer power-orthogonal and, in general, do not form a complete set. From a numerical perspective, the problem of completeness is circumvented by noticing that simulation domains are finite and therefore fall in a special case, where completeness is recovered.

We tackle the problem of orthogonality by investigating the following generalized eigenproblem (EVP) and its transpose:

\begin{align}
        \hat{A}\vec\Psi^R_k = \lambda^R_k \hat{B} \vec \Psi^R_k,\\
        \hat{A}^T\vec\Psi^L_k = \lambda_k^L \hat{B}^T \vec\Psi^L_k,
        \label{EVP}
\end{align}
where $\hat A$ and $\hat B$ are operators, the superscript $``T"$ denotes the transposition and $\lambda^R_k$ ($\lambda^L_k$) is the $k_\text{th}$ right (left) generalized eigenvalue. The transposed EVP is sometimes called adjoint EVP (superscript $\dagger$)~\cite{Siegman1986Lasers}, while its eigenvectors are called left ($\vec\Psi^L_k$) eigenvectors. In the same spirit, eigenvectors of the EVP are the right ($\vec\Psi^R_k$) eigenvectors~\cite{Seyranian2003MultiparameterApplications, Pick2017GeneralPoints}. The EVP and transposed EVP share the same spectrum. At this point we do not restrict our discussion to any specific class of operators as long as no singularities, e.g. exceptional points, are present.

We now recall the definition of an adjoint (transposed)~\footnote{Not to be mistaken with the hermitian adjoint, or transposed and complex conjugated operator, which is only valid if the choice of inner product is the same as in quantum mechanics (bra-ket).} operator. Given a pseudo-inner product~\cite{Mrozowski1997GuidedAnalysis}, denoted by $<,>$ acting on a vector Hilbert space (of square-integrable functions), the transposed operator of $\hat \Theta$ is such that:

\begin{equation}
    <\vec\Psi^L_{k'},\hat\Theta\vec\Psi^R_k> = <\hat\Theta^T\vec\Psi^L_{k'},\vec\Psi^R_k>.
    \label{inner_prod}
\end{equation}

Defining:

\begin{equation}
   <\vec\Psi^L_{k'},\vec\Psi^R_k> = \int dV \vec\Psi^L_{k'}\cdot\vec\Psi^R_k,
\end{equation}
one may show that:

\begin{equation}
    \int dV \vec\Psi^L_{k'}\cdot\hat{A}\vec\Psi^R_k = \int dV \hat{A}^T\vec\Psi^L_{k'}\cdot\vec\Psi^R_k \implies \int dV (\lambda^R_{k}\vec\Psi^L_{k'}\cdot\hat{B}\vec\Psi^R_k-\lambda_{k'}^L\hat{B}^T\vec\Psi^L_{k'}\cdot\vec\Psi^R_k) = 0.
    \label{step1}
\end{equation}

From the above definitions, it is straightforward to prove that if the operator $\hat \Theta$ is a matrix of complex numbers, the transposed operator is obtained by $\Theta_{ij} = \Theta^T_{ji}$; while in the case of differential operators, the transpose can be found using integration by parts.

Taking into account the fact that the EVP and its transpose share the same spectrum, Eq.~\ref{step1} may be simplified to:

\begin{equation}
   (\lambda^R_{k}-\lambda^R_{k'}) \int dV \vec\Psi^L_{k'}\cdot\hat{B}\vec\Psi^R_k = 0,
    \label{step2}
\end{equation} 
which is the statement of bi-orthogonality.

\subsection{Standing-wave resonators}

The task is to find the transpose of the operators of $\bm {\varepsilon}(\vec{r})$ and $\nabla \times \nabla \times$. This will allow us to relate left and right eigenvectors. In electromagnetism, the operator $\hat{B}$ translates into the permittivity tensor $\bm{\varepsilon}$, which in reciprocal media obeys $\bm{\varepsilon} = \bm{\varepsilon}^T$. Also, for standing-wave resonators, $\hat \Theta =\nabla \times \nabla \times$ is symmetric, i.e.  $\hat \Theta = \hat \Theta^T$, as can be shown through integration by parts of this operator acting on any of the entries of Eq.~\ref{inner_prod}~\footnote{Note that in this step we neglect surface terms. This is rigorous since in our simulation domains are limited by perfect electical conductors (PEC) or perfect magnetic conductors (PMC) and therefore field components tangent to the domain's limiting boundary are zero.}.  In other words, the EVP and transposed EVP are the same, therefore the left and right eigenvectors are the same:

\begin{equation}
    \vec E^L_{k} =  \vec E^R_{k},
\end{equation}
Finally, the orthogonality relations become:

\begin{equation}
   \bigg(\frac{\omega_k^2}{c^2}-\frac{\omega_{k'}^2}{c^2}\bigg) \int dV \vec E_{k'}\cdot\bm{\varepsilon}\vec E_{k} = 0,
    \label{step3}
\end{equation}
where we dropped all the superscripts and all fields are right eigenvectors of the EVP.

\subsection{Traveling-wave resonators}

An important case in photonics regards traveling-wave resonators. If the system is periodic, the eigenmodes are Bloch-Floquet type functions:

\begin{equation}
    \vec{\Psi}_{\vec{\beta}}(\vec{r}) = \vec{\psi}_{\vec{\beta}}(\vec{r})e^{i\vec{\beta}\cdot\vec{r}},
\end{equation}
where $\vec{\psi}_{\vec{\beta}}(\vec{r})$ fulfills the periodicity condition for a lattice parameter $\vec{R}$, i.e $\vec{\psi}_{\vec{\beta}}(\vec{r}+ \vec{R}) = \vec{\psi}_{\vec{\beta}}(\vec{r})$ and $\vec{\beta}$ is a vector of the reciprocal lattice, or, in photonics, the propagation vector.
This \textit{ansatz} allows us to modify the electromagnetic eigenvalue problem in terms of the propagation vector $\vec{\beta}$ and the periodic function $\vec{E}^R_{\vec{\beta}}(\vec{r})$:

\begin{equation}
     (\nabla + i\vec{\beta})\times(\nabla + i\vec{\beta})\times \vec{E}^R_{\vec{\beta}}(\vec{r}) = \bm{\varepsilon}(\vec{r})\left(\frac{\omega_{\vec{\beta}}}{c}\right)^{2}\vec{E}^R_{\vec{\beta}}(\vec{r}).
\end{equation}

The transpose of the equation above may again be obtained through integration by parts:

\begin{equation}
    (\nabla - i\vec{\beta}) \times (\nabla - i\vec{\beta}) \times \vec{E}^L_{\vec{\beta}}(\vec{r}) = \left(\frac{\omega_{\vec{\beta}}}{c}\right)^{2}\bm{\varepsilon}(\vec{r}) \vec{E}^L_{\vec{\beta}}(\vec{r}).
\end{equation}
where it finally becomes clear that if the system is periodic, the transpose equation is equivalent to that of counter-propagating modes, therefore obtaining:

\begin{equation}
    \vec{E}^L_{\vec{\beta}} =  \vec{E}^R_{-\vec{\beta}}.
\end{equation}

\subsection{Dispersive media}

When dispersive dielectric materials are considered, the orthogonality relations and perturbation theory results are necessarily modified. The wave equation for dispersive media in frequency domain reads:

\begin{equation}
    \nabla \times \nabla \times \vec{E}^R_{\vec \beta,\omega}(\vec{r})=\left(\frac{\omega_{\vec \beta}}{c}\right)^{2} \bm{\varepsilon}(\vec{r},\omega_{\vec\beta}) \vec{E}^R_{\vec \beta,\omega}(\vec{r}).
    \label{eigenvalue_disp}
\end{equation}
where the presence of sub-indexes $\vec{\beta}$ and $\omega$ accomodates the case of traveling-wave resonators, while the results for the standing-wave case are recovered by simply dropping $\vec{\beta}$.
    Considering a left eigenmode $\vec{E}^L_{\vec \beta',\omega'}$, with frequency $\omega'$, and projecting both sides of the equation above on it, one gets:

\begin{equation}
    \int dV \vec{E}^L_{\vec \beta',\omega'}(\vec{r})\cdot \nabla \times \nabla \times \vec{E}^R_{\vec \beta,\omega}(\vec{r})=\left(\frac{\omega_{\vec \beta}}{c}\right)^{2} \int dV \vec{E}^L_{\vec \beta',\omega'}(\vec{r})\cdot\bm{\varepsilon}(\vec{r},\omega_{\vec \beta}) \vec{E}^R_{\vec \beta,\omega}(\vec{r}).
\end{equation}
Integrating the LHS by parts while neglecting surface terms, yields:

\begin{equation}
  \int dV \vec{E}^L_{\vec \beta',\omega'}(\vec{r})\cdot(\omega^{' 2}_{\vec \beta'}\bm{\varepsilon}(\vec{r},\omega'_{\vec \beta'})- \omega_{\vec \beta}^{2}\bm{\varepsilon}(\vec{r},\omega_{\vec \beta})) \vec{E}^R_{\vec \beta,\omega}(\vec{r})=0.
  \label{orth_derivation}
\end{equation}
This is a statement of orthogonality relations in dispersive media. As discussed before, in traveling-wave resonators $\vec{E}^L_{\vec{\beta}',\omega'} = \vec{E}^R_{-\vec{\beta}',\omega'}$.
\section{Perturbation theory}
For brevity, we state only the results for the dispersive/standing-wave case, as the non-dispersive is directly obtained from it. With the orthogonality relations derived in mind, we may come back to Eq.~\ref{eigenvalue_disp} and derive a novel perturbation series expansion. Our goal is to find an expression to the shift in frequency $\Delta\omega$ of a given eigenmode $\vec{E}^R_{\omega_0}(\vec{r})$ due to a modification in the permittivity of the system $\Delta\bm{\varepsilon}$. We make the following substitutions:

 \begin{gather}
     \bm{\varepsilon}(\vec{r},\omega) \rightarrow \bm{\varepsilon}(\vec{r},\omega_0) + \eta \frac{\partial\bm{\varepsilon}(\vec{r},\omega)}{\partial\omega}\bigg|_{\omega_0}\Delta\omega + \eta\Delta\bm{\varepsilon}(\vec{r},\omega_0),\\
   \omega \rightarrow \omega_0 +\eta\Delta\omega,\\ 
   \vec{E}^R_{\omega}(\vec{r}) \rightarrow \vec{E}^R_{\omega_0}(\vec{r}) + \eta\Delta\vec{E}^R_{\omega_0}(\vec{r}),
   \label{rules}
 \end{gather}
where $\eta$ is the perturbation parameter and $\omega_0$ is the unperturbed frequency. Importantly, the term $\frac{\partial\bm{\varepsilon}(\vec{r},\omega)}{\partial\omega}$ accounts for the change in permittivity due to dispersion.

Assuming that $\Delta\vec{E}^R_{\omega_0}(\vec{r})$ can be expanded in the quasinormal-modes of the system, i.e. $\Delta\vec{E}^R_{\omega_0}(\vec{r}) = \sum_{\omega'} c_{\omega'}\vec{E}^R_{\omega'}(\vec{r})$, we may project the left and right sides of equation Eq.~\ref{eigenvalue_disp} (under the rules in Eq.~\ref{rules}) onto the left eigenvector $\vec{E}_{\omega_0}^L(\vec{r})$. The first correction (order $\eta$) to the eigenvalue is given by:

\begin{equation}
    \Delta \omega = -\frac{\omega_{0}}{2}\frac{\int dV \vec{E}^L_{\omega_{0}}(\vec{r})\cdot\Delta\varepsilon(\vec{r})\vec E^R_{\omega_0}(\vec{r})}{\int dV \vec{E}^L_{\omega_0}(\vec{r})\cdot \big(\bm\varepsilon(\vec{r},\omega_{0})+\frac{\omega_{0}}{2}\frac{\partial \bm \varepsilon(\vec{r},\omega_{0})}{\partial\omega}\big)\vec E^R_{\omega_0}(\vec{r})}.
    \label{pert_theory}
\end{equation}

The analogous expression for the traveling-wave case is given in the main text and can be obtained by projecting Eq.~\ref{eigenvalue_disp} on the counter-propagating pair of $\vec{E}_{\omega_0}$. An important remark must be made here: since the system is non-Hermitian, $\omega_0$ and the volume integrals in Eq.~\ref{pert_theory} are, in general, complex numbers, and thus the imaginary part of $\Delta\omega$ can be related to the modification in the losses of the system.

Normal-mode results are obtained by simply replacing $\vec{E}^L$ and $\vec{E}^R$ with complex conjugate pairs of fields, i.e. $\vec{E}^L= [\vec{E}^R]^*$, where $*$ denotes complex conjugation.
We emphasize that in all normal-mode calculations in the main text, the PML region is excluded from the domains of integration of Eq.~\ref{pert_theory}. This is necessary since the conjugated inner product is not an invariant through PML-type coordinate transforms, in stark contrast with the bi-orthogonal product. In this case, the dissipative coupling arises from the imaginary part of $\omega_0$, unless a complex permittivity (absorber or gain medium) is present in the system. 

We may now proceed and derive expressions for the optomechanical coupling in microresonators, with little importance given to the hermiticity of operators and under the hypothesis of the medium's reciprocity. In the optomechanical case $\Delta\varepsilon(\vec{r}) \propto x$, where $x$ is the mechanical displacement. The first order correction to the optical losses due to a change in the system's permittivity is:

\begin{equation}
    \Delta\kappa_i(x) = 2 \text{Im}\{\Delta \omega( x)\},
\end{equation}
where we assumed a harmonic dependency of the kind $\text{exp}(j\omega t)$. The rate of variation in dissipation due to deformations is:

\begin{equation}
    G_\kappa =  2 \frac{\partial \text{Im}\{\omega\}}{\partial x},
\end{equation}
where we defined $G_\kappa =   \frac{\partial \kappa_i}{\partial x}$. On the dispersive side:

\begin{equation}
    G_\omega =  - \frac{\partial \text{Re}\{\omega\}}{\partial x},
\end{equation}
where the negative sign is adopted out of convention and is accounted for in the equations for the optical mode evolution. The zero-point couplings are defined as  $g_\omega = x_\text{zpf} G_\omega$ and $g_\kappa = x_\text{zpf} G_\kappa$, where $x_\mathrm{zpf}=\sqrt{\hbar/(2m_\mathrm{eff}\Omega_m)}$ is the zero-point fluctuation for the mechanical displacement.

\subsection{Mesoscale optomechanics}

As reported in the main text, in mesoscopic optomechanical systems, the photon-phonon interaction is typically described -- at the level of perturbation theory -- from two main mechanisms, namely: boundary movement and photoelasticity. Although those mechanisms are widely known to the integrated optomechanics community, here we briefly comment how those can be incorporated into the description above.

The change in the dielectric properties of a material due to internal elastic deformations is modeled through the following equation:

\begin{equation}
    \Delta \bm \varepsilon = - \bm \varepsilon^2 \bm p \mkern1mu{:} \bm S,
    \label{eq:pe}
\end{equation}
where $\bm p$ is the photoelastic tensor, $\bm S$ is the elastic strain tensor, which carries the mechanical displacement dependency that is required by the optomechanical coupling, and the $``\mkern1mu{:}"$ operation denotes tensor contraction. $\bm S$ is defined as $ \bm S = \hat\nabla \vec{U}(\vec{r},t)$, where $\hat\nabla = \frac{1}{2}(\nabla+\nabla^T)$ is the symmetrized gradient operation.  Mechanical mode analysis allows us to decompose $\vec{U}$ into acoustic eigenmodes as $\vec{U}(\vec{r},t) = \sum_k \vec{u}_k(\vec{r})x_k(t)$. The latter expression, along with Eqs.~\ref{eq:pe} and~\ref{pert_theory} yields:

\begin{equation}
    \sum_k x_k (t) G_{\text{PE},k} = \frac{\omega_{0}}{2}\frac{\int dV  \vec{E}^L_{\omega_{0}}(\vec{r})\cdot \sum_k\big( \bm \varepsilon^2 \bm p \mkern1mu{:} \hat\nabla \vec{u}_k x_k(t)\big) \vec E^R_{\omega_0}(\vec{r})}{\int dV \vec{E}^L_{\omega_0}(\vec{r})\cdot \big(\bm\varepsilon(\vec{r},\omega_{0})+\frac{\omega_{0}}{2}\frac{\partial \bm \varepsilon(\vec{r},\omega_{0})}{\partial\omega}\big)\vec E^R_{\omega_0}(\vec{r})}.
    \label{eq:pert_theory_PE}
\end{equation}
Where we defined the photoelastic contribution to the optomechanical coupling of the $k$-th mechanical mode as $G_{\text{PE},k} = \frac{\partial\omega}{\partial x_k}\big|_\text{PE}$.

The moving boundary (MB) contribution is slightly non-trivial since it involves electric fields on the interface between two dielectrics (guiding and exterior). Writing Eq.~\ref{pert_theory} in terms of continuous field components has been accomplished by considering the surface as a limit of a smoothing function, as described in detail in Ref.~\cite{Johnson2002PerturbationBoundaries}. In that case, the MB contribution is written as:

\begin{equation}
    \sum_k x_k (t) G_{\text{MB},k}   = -\frac{\omega_{0}}{2}\frac{\int d\vec{A}{\cdot}\big(\sum_k \vec{u}_k(\vec{r})x_k(t)\big)\left( \vec{E}^{L}_{\omega_0,\parallel}(\vec{r}){\cdot}\Delta\bm\varepsilon\vec E^{R}_{\omega_0,\parallel}(\vec{r})-\vec{D}^{L}_{\omega_0,\perp}(\vec{r}){\cdot}\Delta(\bm\varepsilon^{-1})\vec D^{R}_{\omega_0,\perp}(\vec{r}) \right)}{\int dV \vec{E}^L_{\omega_0}(\vec{r}){\cdot}\big(\bm\varepsilon(\vec{r},\omega_0)+\frac{\omega_{0}}{2}\frac{\partial \bm \varepsilon(\vec{r},\omega_{0})}{\partial \omega}\big)\vec E^{R}_{\omega_0}(\vec{r})},
\end{equation}
where modal decomposition of the total surface displacement has been applied, $\vec{D}_{\omega_0}^{R(L)}$ denotes the electric displacement field related to the right (left) electric field, $\perp$ and $\parallel$ denote the perpendicular and parallel field components at the mechanical resonator's surface, and  $\Delta\bm\varepsilon=\bm\varepsilon_1-\bm\varepsilon_2$, $\Delta(\bm\varepsilon^{-1})~=~(\bm\varepsilon_1)^{-1}-(\bm\varepsilon_2)^{-1}$, are related to the permittivities of the guiding ($\bm\varepsilon_1$) and surrounding ($\bm\varepsilon_2$) media. The moving-boundary contribution to the optomechanical coupling of the $k$-th mechanical mode was defined as $G_{\text{MB},k} = \frac{\partial\omega}{\partial x_k}\big|_\text{MB}$.

\subsection{Molecular Optomechanics}

We now treat the case of a point dipole interaction with a single optical mode. The dipole has a mechanical displacement-dependent polarizability given by $\delta(\vec r - \vec{r}_0)\alpha(\{x_k\}) =  \delta(\vec r - \vec{r}_0)(\alpha_0 + \sum_k \frac{\partial \alpha}{\partial x_k}x_k)$, where $\delta(\vec r)$ is the Dirac delta function and $\vec{r}_0$ is the position of the molecule. We note that in  Eq.~\ref{pert_theory}, $\Delta\varepsilon(\vec{r})\vec E^R_{\omega_0}(\vec{r})$ may be straightforwardly associated with a modification in the polarization in the interior of the dielectric. This interpretation allows us to write $\Delta\varepsilon(\vec{r})\vec E^R_{\omega_0}(\vec{r}) = \delta(\vec r - \vec{r}_0)(\sum_k \frac{\partial \alpha}{\partial x_k}x_k) \vec E^R_{\omega_0}(\vec{r})$, yielding:

\begin{equation}
    \sum_k x_k (t) G_k = -\frac{\omega_0}{2} \bigg(\sum_k \frac{\partial \alpha}{\partial x_k}x_k(t)\bigg) \frac{ \vec{E}_{\omega_0}^L(\vec{r}_\text{0})\cdot\vec{E}_{\omega_0}^R(\vec{r}_0)}{\int dV \vec{E}^L_{\omega_0}(\vec{r}){\cdot}\big(\bm\varepsilon(\vec{r},\omega_0)+\frac{\omega_{0}}{2}\frac{\partial \bm \varepsilon(\vec{r},\omega_{0})}{\partial \omega}\big)\vec E^{R}_{\omega_0}(\vec{r})} = -\frac{\omega_{0}}{2 \varepsilon(\vec{r}_{\text{max}}) V_{m}} \sum_k \frac{\partial \alpha}{\partial x_k}x_k(t),
\end{equation}
where $V_m$ denotes the complex modal volume of the optical mode in question. In accordance with Ref.~\cite{Roelli2016MolecularScatteringb}, $\frac{\partial \alpha}{\partial x_k}$ may be written in terms of mass-normalized coordinates (typically denominated $Q_k = \sqrt{m_{\text{eff},k}} x_k$) and the Raman activity $R_k = (\frac{\partial \alpha}{\partial Q_k})^2$. The contribution arising from each mechanical mode is given by evaluating $G_k = \frac{\partial\omega}{\partial x_k}$. In our simulations, the dipole is taken to be optimally aligned/placed with respect to the electric field in the gap between the nanoparticle and mirror, i.e. close to the point of maximum electric field amplitude.

\section{Analytic example}
 QNMs (or the eigenmodes of an open, lossy system) are known to yield divergent real modal volumes (i.e. when power-orthogonality is used), while complex modal volumes (based on bi-orthogonality) remain finite. As shown in~\cite{Sauvan2013TheoryResonators}, the bi-orthogonal product displayed here is an invariant through complex coordinate stretching transforms such as in perfectly matched layers (PML), hence analytic results are expected to be robust to PML position changes. Furthermore, all volume integrations described before must also be carried out in the PML domain. This gives us a very practical way to compute those modal volumes numerically. 

In this section, we use an analytical model to show the convergence of the bi-orthogonal product in open systems and also for insight on how to correctly handle numerical implementations of the PML and complex modal volumes. We consider an infinitely long nanocylinder,  (GaAs, radius $R=\SI{250}{nm}$), which admits whispering-gallery type traveling waves, surrounded by an absorbing layer (PML) also implemented analytically. The PML and cylinder are separated by a gap that will be varied to show the invariance of the bi-orthogonal product with respect to the PML position. For that purpose, we consider perturbations induced by boundary movement (increase in $R$) in the cylinder and calculate (using the moving boundary optomechanical coupling shown in the main text) the shift in frequency of the resonator.

\subsection{Optical modes}

The invariance in $z$ of our system allows to uncouple the $\hat z$ fields from the $\hat r$ and $\hat \phi$ fields, where we adopted cylindrical coordinates in Eq.~\ref{eigenvalue}. An arbitrary solution can be obtained by superposition of two field polarizations - $\vec{E}_z \neq 0, \vec{H}_z = 0$ (TM) and $\vec{E}_z = 0, \vec{H}_z \neq 0$ (TE). Here, we consider only TE modes. 

Using the ansatz $\vec{H}(r) = \Psi_z(r) \text{exp}(-jm\phi) \hat{z}$, the solutions to the $\hat z$ fields are given by the Bessel functions of first and second kinds:

\begin{equation}
    \Psi_z = A J_m\bigg(\frac{n \omega c}{r}\bigg) + B Y_m\bigg(\frac{n \omega c}{r}\bigg).
    \label{zsol}
\end{equation}
In the case of TE modes, only the $\hat{z}$ component of the magnetic field is non-zero, therefore the electric field may be obtained through Maxwell's relations as:

\begin{equation}
    \vec{E}(\vec{r})=-\frac{j}{\omega\varepsilon}\nabla \times \vec{H}(\vec{r}).
\end{equation}

Counter-propagating modes ($m\rightarrow -m$ in Eq. \ref{zsol}) are easily obtained from the solutions of the original problem by flipping signal of the $\hat r$ (and $\hat z$) components of the electric field. This will be useful when using the pseudo-inner product defined above.

While this solution applies to all the domains (dielectric, air and PML), certain simplifications can be made: denoting $R$ as the cavity radius, for $r<R$ we may set $B=0$, otherwise the solution would be divergent at $r=0$. The PML is implemented analytically through the map $r\rightarrow R+\mathrm{gap} + (1 - j)\sigma_0\times(r - R-\mathrm{gap})$, where $\sigma_0 = 5$ is chosen. Furthermore, we limit the domain of our solution by placing a perfect electrical conductor at radius $r= R_\mathrm{PEC}$, positioned far enough from the PML/air interface such that convergence, within numerical accuracy, is achieved.

\begin{figure*}[t]
\centering
\includegraphics[width = 16 cm]{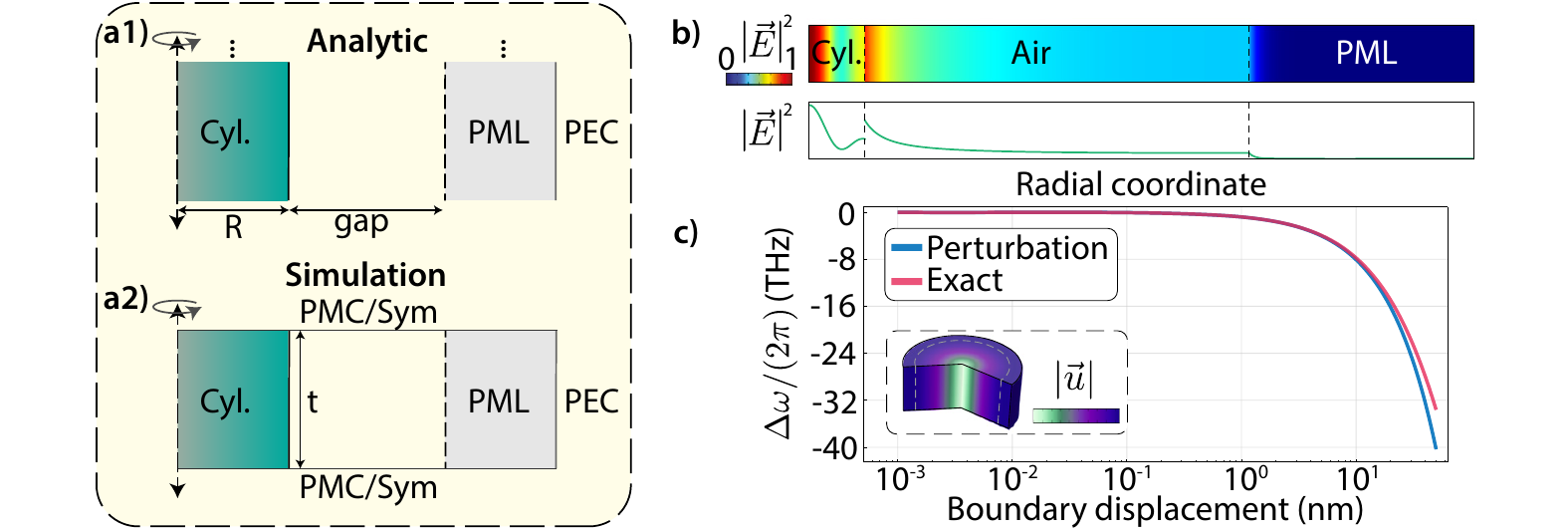}
\caption{\small{ Schematic diagram for the \textbf{a1)} analytic and \textbf{a2)} numerical implementations of the infinite nanocylinder. The perfectly matched layer (PML) is terminated in a perfect electrical conductor (PEC). Perfect magnetic conductor (PMC) boundary conditions are used to emulate an optically infinite cylinder for the TE optical mode under consideration. \textbf{b)} Simulated (upper) and analytic (lower) optical mode profiles, with gap $= \SI{1.7}{\micro\metre}$ and PML size of $\SI{1}{\micro\metre}$. \textbf{c)} Shift in (real) frequency as a function of the increase in radius of the nanocylinder. Inset: Mechanical breathing mode profile.}}
\label{fig:1}
\end{figure*}

For comparison, we implement the same system in COMSOL\textsuperscript{\textregistered}, although a thickness $t =\SI{250}{nm}$ for the cylinder was picked, as shown in Fig.~\ref{fig:1} \textbf{a2)}. This thickness will be important for calculating the mechanical modes supported by the structure, affecting its motional mass and therefore its frequencies. The numerical implementation mimics the invariance in $z$ through the usage of appropriate boundary conditions, e.g. the top and bottom of the nanocylinder are taken to be perfect magnetic conductors. Note that this choice filters out TM modes of the solutions and is therefore consistent with our TE modal analysis. The optical mode profile, in the presence of the PML, is displayed in Fig.~\ref{fig:1} \textbf{b)}, for the case of $\text{gap} = \SI{1.7}{\micro\metre}$, obtained through numerical (upper) and analytic (lower) calculations.

\subsection{Mechanical modes}

The eigenvalue equation for the acoustic normal modes of the structure is given by:

\begin{equation}
    -\rho \Omega_{\mathrm{m}}^{2} \vec{U}(\vec{r})=\nabla \cdot \mathbf{T},
    \label{mech_mode}
\end{equation}
where $\rho$ is the mass density of the material, $\vec{U}(\vec{r})$ is the mechanical mode profile and $\mathbf{T}$ is the stress tensor, which can be found by contraction between the stiffness ($\mathbf{c}$) and strain ($\mathbf{S}$) tensors. We again use the invariance of the system and thus use an \textit{ansatz} $\vec{U}(\vec{r}) = U(r)\hat r$, which will capture purely mechanical breathing modes of the structure. This choice yields only two non-vanishing terms in the strain tensor, namely $S_{1} = \frac{d U}{dr}$ and $S_{2} = \frac{U}{r}$ (in Voigt notation), therefore the solutions to Eq.~\ref{mech_mode} are:

\begin{equation}
    U(r)=A J_{1}\left(\frac{\Omega_{\mathrm{m}} r}{v_{\mathrm{L}}}\right) + B Y_{1}\left(\frac{\Omega_{\mathrm{m}} r}{v_{\mathrm{L}}}\right),
\end{equation}
where $v_L = \sqrt{c_{11}/\rho}$ . Physical solutions must have $B=0$ while $A$ is chosen based on normalization requirements. The mechanical frequency $\Omega_m$ is found by imposing free boundary-conditions ($\mathbf{T}\cdot \hat r = 0$), yielding the following transcendental equation:

\begin{equation}
    \frac{c_{11}}{2v_L}\bigg[J_0\bigg(\frac{\Omega_m R}{v_L}\bigg) - J_2\bigg(\frac{\Omega_m R}{v_L}\bigg)\bigg] + 
\frac{c_{12}}{\Omega_m R}J_1\bigg(\frac{\Omega_m R}{v_L}\bigg) = 0,
\end{equation}
which can be self-consistently solved. 

Numerically, the mechanical mode is computed by considering symmetry boundary conditions ($\hat n \cdot \vec{U} = 0$) on the top and bottom of the nanocylinder, imposing the absence of any mechanical deformations in the $z$ direction, while the $\phi$ dependency is accounted for by choosing the azimuthal number $m=0$. The values found for the mechanical frequencies are $\Omega_m = \SI{6.418}{\giga\hertz}$ (numerical) and $\Omega_m = \SI{6.417}{\giga\hertz}$ (analytic), displaying excellent mutual agreement.

\begin{figure*}[t]
\centering
\includegraphics[width = 16 cm]{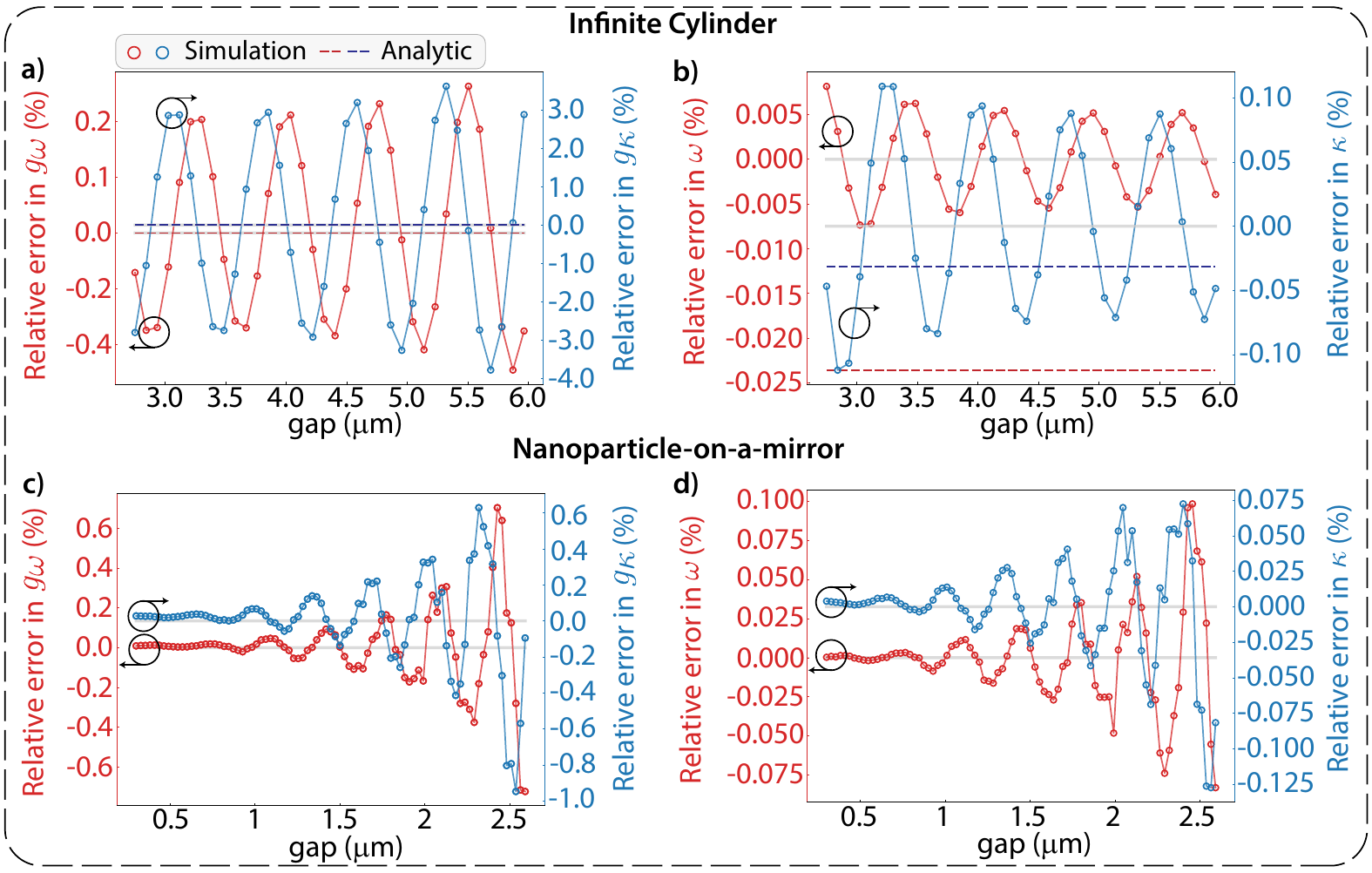}
\caption{\small{\textbf{a)} Relative errors in the optomechanical couplings $g_\omega$, $g_\kappa$ and \textbf{b)} in the (real) frequencies $\omega$ and optical linewidths $\kappa$  as a function of the gap size for an infinite nanocylinder. Both simulated and analytic calculations are shown and validate our FEM results within negligible errors. Averages over all simulated points are depicted by horizontal grey lines. \textbf{c)} and \textbf{d)} display the same analysis for the NPoM (Nanoparticle-on-a-mirror) in the main text, but in the absence of analytic results, which are not available for the geometry considered.}}
\label{fig:2}
\end{figure*}

\subsection{Domain of validity}
We finally explore the extent of validity of the bi-orthogonal perturbation theory. Our aim here is to demonstrate the accuracy and also show how large a deformation has to be (relatively to the size of the structure), in order to significantly deviate from the first order perturbation theory expansion. We consider only boundary movement (increase in $R$), which allows for completely analytical results, in contrast to photoelasticity, where the non-trivial dielectric function imposed by the mechanical strain renders an electromagnetic eigenvalue equation that can only be solved numerically. Also, an increase in $R$ is exactly emulated by the mechanical breathing mode (since displacements are purely radial, as shown in the inset of Fig.~\ref{fig:1} \textbf{c)}), yielding direct comparison between exact and perturbative analytic calculations.

We compute the frequency shift as a function of the deformation of the nanocylinder boundary. As shown in Fig.~\ref{fig:1} \textbf{c)}, the exact and perturbative calculated shifts are in good agreement up to displacements of about $\SI{10}{\nano\metre}$, about $4\%$ of the nanocylinder radius. This puts optomechanical calculations safely within the validity region, where displacements are typically $\ll 0.1\%$ of the device's relevant size.

\subsection{Invariance of the bi-orthogonal product with the PML position}

We now turn our attention to investigating the invariance of the integral in Eq.~\ref{orth_derivation} with respect to the size of the air domain.  This will give us insight on how to correctly position the PML in our numerical study. We start by comparing the results for the zero-point optomechanical couplings in the numerical and analytic models for the nanocylinder ($\omega/(2\pi) \approx \SI{200}{\tera\hertz}$, $\kappa/(2\pi) \approx \SI{27}{\tera\hertz}$, $g_\omega/(2\pi) \approx \SI{2}{\mega\hertz} $, $g_\kappa/(2\pi) = -\SI{260}{\kilo\hertz}$). We sweep over the gap distance (depicted in Fig.~\ref{fig:1} \textbf{a1)} and \textbf{a2)}) between the dielectric's and PML boundaries. For each gap, the optical mode,   $g_\kappa$ and $g_\omega$ are evaluated. The result is displayed in Fig.~\ref{fig:2} \textbf{a)} where simulated points are interpolated as a guide for the eye. The $y$-axis is set to depict relative deviations from the values obtained through averaging $g_\kappa$ and $g_\omega$ for the various gap sizes considered; average values clearly yield 0 error and are represented by grey horizontal lines. Small oscillations (error $<4\%$) in the simulation results are observed, while constant values are obtained in the analytic calculations.  The oscillations in $g_\kappa$ ($g_\omega$) seem to be inherent to the FEM solvers used; we also verify $g_\kappa$ ($g_\omega$) to display the same periodicity of the imaginary (real) parts of $\vec{E}_{-\vec \beta}\cdot\vec{E}_{\vec \beta}$, which explains the $\frac{\pi}{2}$ phase difference observed between the blue and red curves. The increasing oscillation amplitudes are related to the exponential growth of the QNM fields as a function of the distance to the PML domain. Although averaging over many simulations seems to yield completely accurate results, this procedure can be safely ignored in view of the negligibly small errors verified.  Mesh refinement within the computational power available was not observed to yield significantly better results than those displayed in here. The same features are present if the frequency and linewidth of the optical modes are considered. Results are displayed in Fig.~\ref{fig:2} \textbf{b)} and notably, the relative error in this case (amplitude/average ratio) is much smaller than in the previous case. 

In Figs.~\ref{fig:2} \textbf{c)} and \textbf{d)} an analogous treatment is presented for the case of the nanoparticle-on-a-mirror scheme presented in the main text. Here, the gap length is taken to be the distance between the radius of the nanosphere and the air/PML interface. Negligibly small oscillations in the simulated values are again verified. Lastly, we note that for high-Q resonators, such as the split-beam nanocavity in the text, oscillations were not verified within numerical accuracy.

\section{Dynamical backaction in the NPoM system}
\subsection{Hamiltionian formalism}
The interaction Hamiltonian of an optomechanical system with both dissipative and dispersive interactions is given by: 
$\hat{H} = \hbar\omega_c \hat{a}^\dagger\hat{a} +\hbar\omega_c \hat{b}^\dagger\hat{b} + \hat{H}_\kappa + \hat{H}_{\Gamma_m} + \hat{H}_\mathrm{OM}$ with:
\begin{equation}
    \hat{H}_\text{OM} = -\left[\hbar G_\omega \hat{a}^{\dagger} \hat{a}+\sum_{J=e,i}i \sqrt{\frac{\kappa_J}{2 \pi \rho_J}} \frac{\hbar G_{\kappa_J}}{2\kappa_J} \sum_{q}\left(\hat{a}^{\dagger} \hat{c}_{J,q}-\hat{c}_{J,q}^{\dagger} \hat{a}\right)\right]\hat x,
    \label{Hamiltonian}
\end{equation}
where $\hat a \, (\hat a^\dagger)$ and $\hat{c}_{J,q} \, (\hat{c}_{J,q}^\dagger)$ are the bosonic annihilation (creation) operators for the cavity and bath optical modes respectively. The subscript $J$ in the bath operators splits their contribution in the external $\kappa_e$ and internal $\kappa_i$ decay channels. The difference arises from the fact that the external channel is coherently driven by an external laser. In that sense, the total cavity decay rate is written $\kappa = \kappa_e+\kappa_i$. Lastly, $\hat x$ is the mechanical position operator, and $\rho_J$ denotes the density of states of the optical bath, treated as a constant for the relevant frequencies. Equation~\ref{Hamiltonian}, along with Hamiltonians for the isolated optical/acoustic modes and their respective damping, allow us to formulate the dynamics of the system at the level of quantum Langevin equations~\cite{Elste2009QuantumNanomechanics, Weiss2013QuantumSystems}.

The linearized optical force operator is directly evaluated from $\hat{H}_\text{OM}$ as $\hat F = - \frac{\partial \hat H_\text{OM}}{\partial \hat x}$:

\begin{equation}
\begin{aligned}
\hat{F} x_{\text{zpf}} &=\hbar\tilde{A} \kappa\left(\bar{a}^{*} \hat{d}+\bar{a} \hat{d}^{\dagger}\right) \\
&+i \frac{\hbar\tilde{B}_e}{2} \sqrt{\kappa_{\text {e}}}\left[\bar{a}^{*} \hat{d}_{\text {in}}^{\text{e}}-\left(\hat{d}_{\text {in}}^{\text{e}}\right)^{\dagger} \bar{a}\right] 
+i\frac{\hbar\tilde{B}_e}{2}\left[\bar{a}^{*}\left(i \Delta+\frac{\kappa}{2}\right) \hat{d}+\bar{a}\left(i \Delta-\frac{\kappa}{2}\right) \hat{d}^{\dagger}\right] \\
&+i \frac{\hbar\tilde{B}_i}{2} \sqrt{\kappa_{i}}\left[\bar{a}^{*} \hat{d}_{\mathrm{in}}^{i}-\left(\hat{d}_{\mathrm{in}}^{i}\right)^{\dagger} \bar{a}\right],
\end{aligned}
\end{equation}
where the optical amplitude has been split into coherent ($\bar a = \langle\hat a\rangle$) and fluctuation ($\hat d$) parts as $\hat a = \bar{a}+ \hat d$; $\tilde{B}_J\kappa_J = G_{\kappa_J}x_\text{zpf}$, $\tilde{A}\kappa = G_\omega x_\text{zpf}$, the $\hat{d}_{\mathrm{in}}^{J}$ denote the input noise through the external ($\hat{d}_{\mathrm{in}}^{e}$) or internal ($\hat{d}_{\mathrm{in}}^{i}$) loss channels and $\Delta=\omega_l-\omega_c$ is the detuning between the laser drive ($\omega_l$) and cavity ($\omega_c$) frequencies. Assuming a zero temperature, Markovian optical bath, noise correlations are: $\left\langle\hat{d}^{J}_{\mathrm{in}}(\omega) (\hat{d}^{J^{\prime}}_{\mathrm{in}}\left(\omega^{\prime})\right)^{\dagger}\right\rangle=2 \pi \delta\left(\omega+\omega^{\prime}\right)\delta_{J,J^{\prime}}$. The first line describes the force arising from the dispersive coupling, while the second and third lines emerge from the external and internal dissipative couplings, respectively. Importantly, the second term in the second line appears due to the coherent drive in the external port. 
In the weak coupling regime, one may compute the cooling/amplification process from the force's noise spectrum $S_\mathrm{FF}(\Omega)$\cite{Elste2009QuantumNanomechanics, Weiss2013QuantumSystems}:

\begin{equation}
\begin{aligned}
    S_\mathrm{FF}(\Omega) &=\bigg(\frac{\hbar\tilde{B}_e}{2 x_\text{zpf}}\bigg)^2 n_c \frac{\kappa_{\text {e}}(\Omega+2 \Delta-2 \tilde{A} \kappa / \tilde{B}_e)^{2}}{(\kappa / 2)^{2}+(\Omega+\Delta)^{2}}+\bigg(\frac{\hbar\tilde{B}_e}{2 x_\text{zpf}}\bigg)^2 n_c \frac{\kappa_{i} [(\Delta-2 \tilde{A} \kappa / \tilde{B}_e)^{2}+\kappa^{2} / 4]}{(\kappa / 2)^{2}+(\Omega+\Delta)^{2}}\\&+ \bigg(\frac{\hbar\tilde{B}_i}{2 x_\text{zpf}}\bigg)^2 n_c \frac{\kappa_{i}[(\Omega+\Delta-2 \tilde{A} \kappa / \tilde{B}_i)^{2}+\kappa^{2} / 4]}{(\kappa / 2)^{2}+(\Omega+\Delta)^{2}} +\bigg(\frac{\hbar\tilde{B}_i}{2 x_\text{zpf}}\bigg)^2 n_c \frac{\kappa_{\text {e}}(2 \tilde{A} \kappa / \tilde{B}_i)^{2}}{(\kappa / 2)^{2}+(\Omega+\Delta)^{2}}\\
    &- \frac{n_c}{x_\text{zpf}^2}\frac{ \hbar^2 \tilde{A}^2 \kappa^3}{(\kappa / 2)^{2}+(\Omega+\Delta)^{2}}+ \frac{\hbar^{2} \tilde{B}_{e} \tilde{B}_{i}}{2 x_\text{zpf}^2}n_c\frac{\kappa_{i}[\Delta(\Delta+\Omega) -\kappa^2/4]}{(\kappa/2)^2+(\Delta+\Omega)^{2}},
\end{aligned}
\end{equation}
where $n_c$ is the number of photons inside the cavity. The first  (second) line calculates the interplay between dispersive and external (internal) dissipative couplings in the force spectrum. In the first term of the third line we subtract the purely dispersive contribution to $S_\mathrm{FF}(\Omega)$ since it has been accounted for twice in the first two lines. The last term in the third line describes the interference between the two dissipative couplings.

 Fermi’s Golden Rule for transitions between states with $n$ and $n\pm 1$ phonons results in an optically induced mechanical damping $\delta\Gamma=\Gamma_{\downarrow}-\Gamma_{\uparrow}$ ($\Gamma_{\uparrow(\downarrow)}=\frac{x_\mathrm{zpf}^2}{\hbar^2}
 S_\mathrm{FF}(\mp\Omega_m)$), from which the results in the main text are recovered. Importantly, under the approximations used in the main text, and justified in the subsection below, i.e. $\tilde{B}_e/\tilde{B}_\text{rad}\approx 1$, $\delta\Gamma/n_c$ becomes independent of $\kappa_e$ and $\kappa_i$ and can be fully expressed in terms of the total dissipative coupling $g_\kappa$, the absorptive dissipative coupling $g_{\kappa_\text{abs}}$, total losses $\kappa$, and absorptive losses $\kappa_\text{abs}$, i.e. cooling and amplification depend on the correct value of external coupling rate solely through the number of plasmons excited in the resonator.

 \subsection{External and internal dissipative couplings}
 
 As described in the main text, in plasmonic systems such as the NPoM, excitation of QNM's is usually mediated by its radiation field. This allows us to relate the radiative part of the internal dissipative coupling $g_{\kappa_\text{rad}}$ to the external dissipative coupling $g_{\kappa_\text{e}}$. In the following, we demonstrate that $g_{\kappa_\text{e}}/g_{\kappa_\text{rad}} \approx  \kappa_\text{e}/\kappa_\text{rad}$ or $\tilde{B}_e/\tilde{B}_{\text{rad}} \approx 1$, where $\tilde{B}_{\text{rad}} = g_{\kappa_\text{rad}}/\kappa_\text{rad}$. 
 
\begin{figure*}[t]
\centering
\includegraphics[width = 16 cm]{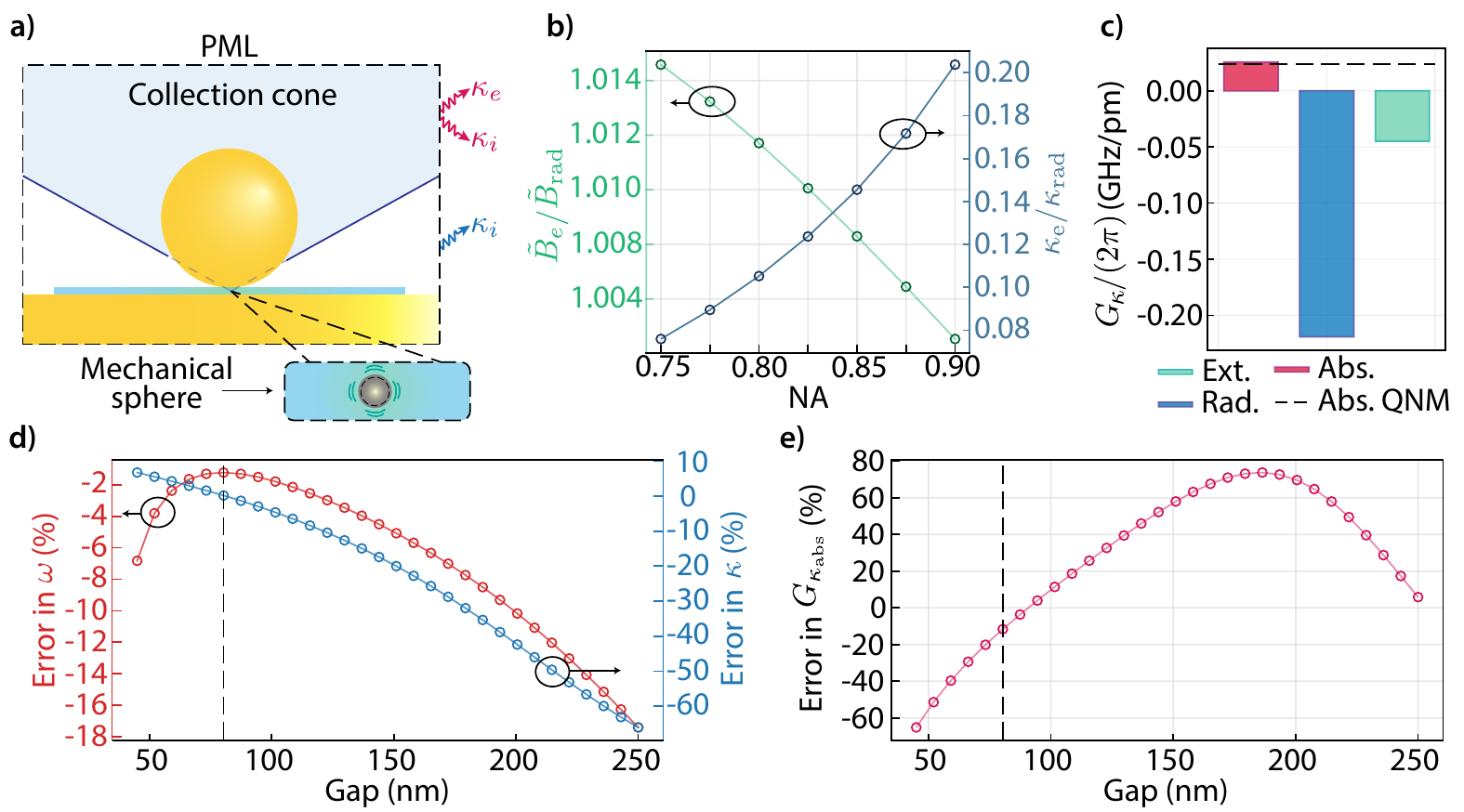}
\caption{\small{\textbf{a)} Illustration of the simulated setup. Radiation into the collection cone is partially used as a coherent excitation channel, with coupling rate $\kappa_e$. The remaining fraction acts as an internal dissipation rate along with radiation outside the cone yielding $\kappa_\text{rad}$. Inset: mechanical mode of the dielectric sphere used in the moving mesh simulations. \textbf{b)} $\tilde{B}_e/\tilde{B}_{\text{rad}}$ and $\kappa_e/\kappa_{\text{rad}}$ as a function of the numerical aperture (NA) used in simulations. The results shown assume $\eta = 0.4$, yielding $\kappa_e/\kappa \approx 0.1$ for NA = $0.9$. \textbf{c)} External (Ext.), absorptive (Abs.), and radiative (Rad.) dissipative coupling rates evaluated through Moving Mesh calculations. Note that the absorptive coupling is evaluated exactly (red bar) and compared to results from QNM perturbation theory (Abs. QNM) in the absence of PMLs in the simulations. \textbf{d)} Errors in $\omega$ (real part of the eigenfrequency) and $\kappa_\text{abs}$ when comparing exact values using PMLs with results in the absence of PMLs where the simulation domain size is varied. \textbf{e)} Errors in $G_{\kappa_\text{abs}}$ when comparing exact values obtained through moving mesh simulations with QNM perturbation theory in the absence of PMLs. In \textbf{d)} and \textbf{e)} the vertical dashed lines represent the gap distance that yields the minimal error in $\omega$.}}
\label{fig:3}
\end{figure*}

 We use FEM simulations of the NPoM scheme coupled to a dielectric ($\varepsilon_r = 3.75$) sub-nanometer sphere with radius $R = \SI{0.3}{\nano\metre}$, which plays the role of the molecule, placed at the center of the gap between the gold nanosphere and mirror. Using moving mesh calculations we deform the dielectric sphere radially and compare the directionality of radiated fields in the deformed and undeformed configurations. This is done by computing the QNM's optical power outflow on the interfaces between the air and PML domains. As shown in Fig.~\ref{fig:3}\textbf{a)}, the power outflow is divided in two different domains: inside (rate $\kappa_\text{col}$) and outside (rate $\kappa_\text{n-col}$) a collection cone related to the numerical aperture (NA) of lenses used in experiments. Due to mode matching, not all collected radiation acts as a driving channel, thus the external coupling $\kappa_e$ is related to an approximately constant fraction, $\eta$, of the coupling rate to the collection cone, i.e. $\kappa_e = \eta \kappa_\text{col}$ and thus $\kappa_\text{rad} = (1-\eta)\kappa_\text{col}+\kappa_\text{n-col}$. Values for $\eta$ are chosen in order to yield coupling efficiencies similar to those obtained in state-of-the-art experiments in NPoMs, i.e. $\kappa_e/\kappa = 10\%$~\cite{Baumberg2019ExtremeGaps}. Note that  using this analysis, we can also monitor the absorption losses given by $\kappa_\text{abs}  =  \kappa-\kappa_\text{col}-\kappa_\text{n-col}$, where $\kappa$ is directly evaluated from the QNM's eigenfrequency.
 
 The analogy between this and the molecular optomechanical cases -- where a point-dipole with displacement-dependent polarizability is considered -- is clear when one notices that the electric field is approximately uniform in the gap region. In this case, we may approximate the electric response of the dielectric sphere by a dipole with polarizability:
 
\begin{equation}
    \alpha = 4 \pi R^3 \varepsilon_{0} \frac{\varepsilon-\varepsilon_{0}}{\varepsilon+2 \varepsilon_{0}},
\end{equation}
 The modification in the polarizability due to a change in the radius is given by $\frac{\partial \alpha}{\partial R} \delta R= 12 \pi R^2 \varepsilon_0 \frac{\varepsilon-\varepsilon_{0}}{\varepsilon+2 \varepsilon_{0}} \delta R$, which closes the link between the molecular and the FEM cases.

In Fig.~\ref{fig:3}\textbf{b)} we evaluate the ratio $\tilde{B}_e/\tilde{B}_{\text{rad}}$ as a function of the numerical aperture considered. We note that the approximation $\tilde{B}_e/\tilde{B}_{\text{rad}} \approx 1$ is backed by our FEM results, in agreement with the discussion made in the main text; for completeness, we also show $\kappa_\text{e}/\kappa_\text{rad}$. The bars in Fig.~\ref{fig:3} \textbf{c)} show the external, radiation, and absorption components of the total dissipative coupling, from which we conclude that the radiative component is dominant. The horizontal dashed line is the estimate of  $G_{\kappa_\text{abs}}$ using QNM perturbation theory in the absence of PMLs. In that case, radiative losses are neglected and any dissipation/dissipative optomechanical coupling necessarily arises from absorption. Since QNMs are inherently divergent if no PMLs are used, this step requires special care. We adopt the following procedure: the size of the domain (here parametrized by the variable ``Gap" -- the largest distance between the center of the gold nanosphere and the PEC used to delimit our solution's domain) is decreased until we minimize the difference in the real part of the eigenfrequency in simulations in the presence and absence of PMLs; this is only accomplished if the fields near the resonator are kept intact while avoiding their exponential increase at large distances. Results obtained during this procedure are shown in Fig.~\ref{fig:3} \textbf{d)} (left \textit{y}-axis). The vertical dashed line shows the size of the gap that yields the smallest error between the real part of the eigenfrequencies of the simulations with and without PMLs. We also compare the obtained values of $\kappa_\text{abs}$ (right \textit{y}-axis); the exact value of $\kappa_\text{abs}$ was calculated using $\kappa_\text{abs}  =  \kappa-\kappa_\text{col}-\kappa_\text{n-col}$ (evaluated with PMLs) and agrees with our estimate without PMLs with errors inferior to $5\%$ . Lastly, in Fig.~\ref{fig:3} \textbf{e)} we compare the QNM perturbation theory estimate of $G_{\kappa_\text{abs}}$ with the value retrieved from moving mesh calculations. The vertical dashed line is the same from Fig.~\ref{fig:3} \textbf{d)} and shows an error of $\approx 10\%$ for the optimal Gap size.
 
\begin{acknowledgments}
\textit{Note}: FEM and scripts files for generating each figure are available at Ref.~\cite{zenodo_nonhermitian}. 
\end{acknowledgments}


%